\documentclass[10pt]{article}
\usepackage{latexsym,amsmath,amssymb,graphics,amscd}
\usepackage{multirow}
\usepackage{soul}
\usepackage{booktabs}
\usepackage{tabularx}
\usepackage[hang,footnotesize]{caption}
\usepackage[pdftex]{graphicx}
\usepackage{graphicx}
\usepackage{color}
\usepackage[pdftex,breaklinks,colorlinks,
  citecolor=blue,
  urlcolor=blue]{hyperref}
\usepackage{tikz}
\usetikzlibrary{decorations.pathreplacing}
\usepackage{xr}
\addtocounter{footnote}{1}
\makeatletter
\@addtoreset{equation}{section}

\makeatother

\newtheorem{theorem}{Theorem}[section]
\newtheorem{definition}[theorem]{Definition}

\newtheorem{remark}[theorem]{Remark}
\newtheorem{proposition}[theorem]{Proposition}

\newtheorem{corollary}[theorem]{Corollary}
\newtheorem{example}[theorem]{Example}

\newcommand{\bfi}{\bfseries\itshape}
\newsavebox{\savepar}

\makeatletter

\begin{document}

\title{\textbf{Quantitative evaluation of the performance of discrete-time reservoir computers in the forecasting, filtering, and reconstruction of  stochastic stationary  signals}}
\author{Lyudmila Grigoryeva$^{1}$, Julie Henriques$^{2, 3}$, and Juan-Pablo Ortega$^{3, \ast}$}
\date{}
\maketitle
 
\begin{abstract}
This paper extends the notion of information processing capacity for non-independent input signals in the context of reservoir computing (RC). The presence of input autocorrelation makes worthwhile the treatment of forecasting and filtering problems for which we explicitly compute this generalized capacity as a function of the reservoir parameter values using a streamlined model. The reservoir model leading to these developments is used to show that, whenever that approximation is valid, this computational paradigm satisfies the so called separation and fading memory properties that are usually associated with good information processing performances. We show that several standard memory, forecasting, and filtering problems that appear in the parametric stochastic time series context can be readily formulated and tackled via RC which, as we show, significantly outperforms standard techniques in some instances. 

\end{abstract}

\bigskip


\makeatletter
\addtocounter{footnote}{1} \footnotetext{%
Department of Mathematics and Statistics. Universit\"at Konstanz. Box 146. D-78457 Konstanz. Germany. {\texttt{Lyudmila.Grigoryeva@uni-konstanz.de} }}
\makeatother
\makeatletter
\addtocounter{footnote}{1} \footnotetext{%
Cegos Deployment. 11, rue Denis Papin. F-25000 Besan\c{c}on. {\texttt{jhenriques@deployment.org} }}
\makeatother

\makeatletter
\addtocounter{footnote}{1} \footnotetext{%
Corresponding author. Centre National de la Recherche Scientifique, Laboratoire de Math\'{e}matiques de Besan\c{c}on, UMR CNRS 6623, Universit\'{e} de Franche-Comt\'{e}, UFR des
Sciences et Techniques. 16, route de Gray. F-25030 Besan\c{c}on cedex.
France. {\texttt{Juan-Pablo.Ortega@univ-fcomte.fr} }}
\makeatother

\section{Introduction}

Reservoir computing is a recent but already well established neural computing paradigm~\cite{jaeger2001, Jaeger04, maass1, maass2, Crook2007, verstraeten, lukosevicius} that has shown a significant potential in overcoming some of the limitations inherent to more standard Turing-type machines. This computation approach, also referred to in the literature as {\bf Echo State Networks} and {\bf Liquid State Machines},  is characterized by a simple and convenient supervised learning scheme, even though its performance presents as a weak side a substantial sensitivity to architecture parameters. This feature explains the development in the literature of various linear and nonlinear memory capacity measures~\cite{Jaeger:2002, White2004, Ganguli2008, Hermans2010, dambre2012}  as well as the study of different signal treatment properties (see~\cite{Yildiz2012, lukosevicius} and references therein) {that are used} to characterize and measure the information processing abilities of these devices in order to be able to optimize them.

We have proposed several contributions in this direction in our previous works~\cite{GHLO2014_capacity, RC3}  in the context of RCs constructed via the sampling of the solutions of a time-delay differential equation. These RCs are usually referred to as time-delay reservoirs (TDRs). More specifically, in~\cite{GHLO2014_capacity} we constructed a simplified model for those specific RCs that allowed us to  provide a functional link between the RC parameters and its performance with respect to a given memory task and which can be used to accurately determine the optimal reservoir architecture by solving a well structured optimization problem. The availability of this tool simplifies enormously the implementation effort and sheds new light on the mechanisms that govern this information processing technique.  This approach was extended in~\cite{RC3} in order to be able to handle multidimensional input signals and real-time multitasking~\cite{maass2}, that is, the simultaneous execution of several memory tasks. Additionally, we used this approach {to} estimate the memory capacity of parallel arrays of reservoir computers. This reservoir architecture had been introduced in~\cite{pesquera2012, GHLO2012}, where it was empirically shown to exhibit various improved robustness properties.

The notion of capacity is defined using independent input signals, which immediately limits its practical functionality in several aspects. Indeed, the use of independent inputs makes empty of content the treatment of forecasting problems. Additionally, most input signals that need to be processed in specific tasks exhibit sizable autocorrelation, which automatically precludes independence. Finally, simple numerical experiments show that optimal reservoir architectures with respect to a given memory task lose that optimality as soon as the input signal ceases to be independent. 

All these facts call for a generalization of the notion of capacity suitable for correlated signals and for techniques to compute it. This is the main goal of this work. More specifically, we use an extension of the RC model introduced in~\cite{GHLO2014_capacity} in order to generalize the  memory capacity formulas that were introduced in that paper to non-independent strictly stationary signals. Moreover, the presence of input autocorrelation makes worthwhile the treatment of forecasting and filtering problems for which we will {\it extend the notion of capacity and that we will explicitly compute as a function of the reservoir parameter values. These results can be readily used in the execution of specific tasks since the expressions that we obtain are written in terms of various statistical features of the input and the teaching signal that can be simply estimated out of the training sample.}

The results in this paper are formulated for general discrete-time RCs that are not necessarily TDRs. We will use the generalization of the model in~\cite{GHLO2014_capacity} to this context  in order to show that, for that approximation, {\it RCs satisfy the so called {fading memory} and {separation properties}} that are typically associated to good information processing performances (see~\cite{Yildiz2012, lukosevicius} and references therein). 

We conclude the paper with a section in which we show that several memory, forecasting, and filtering problems that appear profusely in the context of parametric stochastic time series models can be readily formulated and tackled via RC which, as we show, outperforms in some instances standard techniques in that setup. 

\medskip

\noindent {\bf Notation:} 
column vectors are denoted by bold lower or upper case  symbol like $\mathbf{v}$ or $\mathbf{V}$. We write $\mathbf{v} ^\top $ to indicate the transpose of $\mathbf{v} $. Given a vector $\mathbf{v} \in \mathbb{R}  ^n $, we denote its entries by $v_i$, with $i \in \left\{ 1, \dots, n
\right\} $; we also write $\mathbf{v}=(v _i)_{i \in \left\{ 1, \dots, n\right\} }$. 
The symbols $\mathbf{i} _n$ and $ \mathbf{0} _n $ stand for the vectors of length $n$ consisting of ones and zeros, respectively. 
We denote by $\mathbb{M}_{n ,  m }$ the space of real $n\times m$ matrices with $m, n \in \mathbb{N} $. When $n=m$, we use the symbol $\mathbb{M}  _n $  to refer to the space of square matrices of order 
$n$. Given a matrix $A \in \mathbb{M}  _{n , m} $, we denote its components by $A _{ij} $ and we write $A=(A_{ij})$, with $i \in \left\{ 1, \dots, n\right\} $, $j \in \left\{ 1, \dots m\right\} $.   We write $\mathbb{I} _n $ and $\mathbb{O} _n $ to denote the identity matrix and the zero matrix of dimension $n$, respectively.
We use $\mathbb{S}_n   $ to  indicate the subspace $\mathbb{S}  _n \subset \mathbb{M}  _n $ of symmetric matrices, that is, $\mathbb{S}  _n = \left\{ A \in \mathbb{M}  _n \mid A ^\top = A\right\}$.
Finally, the symbols ${\rm E}[\cdot]$, ${\rm var}(\cdot)$, and ${\rm Cov}(\cdot,\cdot)$ denote the mathematical expectation, the variance, and the covariance, respectively.

\medskip
 
\noindent {\bf Acknowledgments:} We acknowledge partial financial support of the R\'egion de Franche-Comt\'e (Convention 2013C-5493), the ANR ``BIPHOPROC" project (ANR-14-OHRI-0002-02), and Deployment S.L. LG acknowledges financial support from the Faculty for the Future Program of the Schlumberger Foundation.

\section{Reservoir computing with stationary input signals}

We start by spelling out in detail the main dynamical property of the input signals that we consider in this work.
Let $\left\{z(t)\right\}_{t\in \mathbb{Z}}$  be a one-dimensional stochastic time series, that is, for any time $t \in \mathbb{Z}$, the value $z(t) \in \mathbb{R}$ is the realization of a univariate random variable. 

\begin{definition} \label{def strictly stationary}
The time series $\left\{z(t)\right\}_{t\in \mathbb{Z}}$ is said to be \textbf{strictly stationary} if the joint distributions of the multivariate random variables $\left(z(t_1),\dots,z(t_k)\right)^\top$ and $\left(z(t_1+h),\dots,z(t_k+h)\right)^\top$ are the same for all positive integers $k$ and for all $t_1,\dots,t_k,h \in \mathbb{Z}$.
\end{definition}

\begin{definition}
\label{higher order stationary}
Given a time series $\left\{z(t)\right\}_{t\in \mathbb{Z}}$, $r_1,\dots, r_k, k \in \mathbb{N}$ and $t,h_2,\dots,h_k \in \mathbb{Z}$ we define the corresponding \textbf{higher order automoment} $ \mu_{z}^{r_1,\dots,r_k}\left(t,h_2,\dots,h_k\right) $ as

\begin{equation}
 \mu_{z}^{r_1,\dots,r_k}\left(t,h_2,\dots,h_k\right):={\rm E} \left[ z(t)^{r_1}z(t+h_2)^{r_2}\cdots z(t+h_k)^{r_k} \right],
 \label{automoment}
\end{equation}
together with the convention
\begin{equation*}
\mu_{z}^{r_1}(t)={\rm E} \left[ z(t)^{r_1}) \right].
\end{equation*}
\end{definition}
It is straightforward to show that the symbols in~\eqref{automoment} satisfy the following two reduction properties:
\begin{itemize}
\item If $h_i=0$ then: 
\[
\mu_{z}^{r_1,\dots,r_i,\dots,r_k}\left(t,h_2,\ldots,h_i,\dots,h_k\right)= \mu_z^{r_1+r_i,r_2,\dots,r_{i-1},r_{i+1},\dots, r_{k}}\left(t,h_2,\dots,h_{i-1},h_{i+1},\dots,h_k \right).
\]
\item If $h_i=h_j\neq 0$ with $ i<j$ then:
\[
\mu_{z}^{r_1,\ldots, r_k}\left(t,h_2,\dots,h_k\right) = \mu_{z}^{r_1,\ldots, r_{i-1} ,(r_i+r_j) ,r_{i+1},\ldots ,r_{j-1},r_{j+1},\ldots, r_k}\left( t,h_2,\dots,h_i,\dots,h_{j-1},h_{j+1},\dots,h_k\right).
\]
\end{itemize}
 
The following proposition is a direct consequence of Definition \ref{def strictly stationary}.
 
 \begin{proposition}
 Let $\left\{z(t)\right\}_{t\in \mathbb{Z}}$ be a stochastic time series whose higher order automoments exist. If $\left\{z(t)\right\}_{t\in \mathbb{Z}}$ is strictly stationary then its higher order automoments are time-independent. In that case, we replace the notation in \eqref{automoment} by 
 \begin{equation}
  \mu_{z}^{r_1,\dots,r_k}\left(h_2,\dots,h_k\right):={\rm E} \left[ z(t)^{r_1}z(t+h_2)^{r_2}\cdots z(t+h_k)^{r_k} \right] \quad \mbox{for any} \quad t \in \Bbb Z.
 \end{equation}
 \end{proposition}
 
 \begin{remark}
 \label{second order extension}
 \normalfont
 We emphasize that strict stationarity is a significant generalization of the standard notion of stationarity formulated in terms of the time-independence of the order one and two automoments~\cite{Brockwell2002} (mean and autocovariance functions), also referred to as {\bf second order stationarity}. A case in which both notions coincide is when the process  $\left\{z(t)\right\}_{t\in \mathbb{Z}}$ is Gaussian, that is, when for any $t_1,\dots,t_k \in \mathbb{Z}$ and $k \in \mathbb{N}$, the distribution function of $\left( z(t_1),\dots,z(t_k) \right)$ is a multivariate Gaussian. \quad $\blacksquare$
 \end{remark}
 
 \subsection{The RC setup for signal forecasting, filtering, and reconstruction}
 
 \subsubsection{The tasks}
 \label{The tasks}
 In this work we study the performance of reservoir computing in the handling of three different signal processing tasks, namely {\bf forecasting, filtering}, and {\bf reconstruction}, that we now describe in detail. Let $\left\{z(t)\right\}_{t\in \mathbb{Z}}$ and $\left\{y(t)\right\}_{t\in \mathbb{Z}}$ be two one-dimensional stochastic time series that will be called in what follows the {\bf input} and {\bf teaching signals}, respectively. The goal of any machine learning based signal treatment strategy consists of using finite size realizations $\mathbf{z}_T:=\{z(1), \ldots, z(T)\} $ and ${\bf y}_T :=\{y(1), \ldots, y(T)\} $ of $\left\{z(t)\right\}_{t\in \mathbb{Z}}$ and $\left\{y(t)\right\}_{t\in \mathbb{Z}}$, respectively, in order to train a device that is capable of reproducing  out-of-sample realizations  $\mathbf{y}'_{T'}$ of the teaching signal out of a corresponding realization of the input signal {$\mathbf{x}'_{T'} $}. The pairs $(\mathbf{z}_T,\mathbf{y}_T)$ and $(\mathbf{z}'_{T'},\mathbf{y}'_{T'})$ are referred to as {\bf training} and {\bf testing samples}, respectively.
 
If we use the mean square error $ {\rm E} \left[ \left(y(t)-\bar{y}(t)\right)^2\right]$ as a loss function  to measure the difference between the machine output $\bar{y}(t) $ and the actual value $y(t) $ of the teaching signal that we seek to reproduce, a general result shows (see for example Section 4.1 in~\cite{MR1278033}) that this machine learning task amounts to a nonparametric estimation of the conditional expectation ${\rm E} \left[y(t)\left|\mathcal{F}_t \right.\right]$, where $\mathcal{F}_t$ is the information set generated by the input signal up to time $t$, that is, $\mathcal{F}_t =\sigma\left(z(t),z(t-1),\dots\right)$; the symbol $\sigma\left(z(t),z(t-1),\dots\right) $ denotes the sigma-algebra generated by the random variables $\{z(t),z(t-1),\dots\} $. We will distinguish three different, but possibly overlapping situations, that will be illustrated later on in Section~\ref{Examples}:
\begin{description}
\item [(i)] {\bf Forecasting and reconstruction:} in this case the teaching signal is a function, in general nonlinear, of the input signal. More specifically, we define a  \textbf{$(f,h)$-lag forecasting/reconstruction task} as a function $H: \mathbb{R}^{f+h+1}\rightarrow \mathbb{R}$  that is used to generate a one-dimensional signal $y(t)=H\left(z(t+f),\dots,z(t),\dots,z(t-h)\right)$ that depends on the value of the input signal $f$ lags into the future (forecasting part) and $h$ lags into the past (reconstruction part). Examples of this task are presented in Subsections~\ref{Multistep forecast of ARMA temporal aggregates} and~\ref{Multistep forecast of temporally aggregated GARCH volatilities}.
\item [(ii)] {\bf Filtering:} it is a generalization of the previous case in which the input and teaching signal exhibit statistical dependence but do not necessarily have a deterministic functional dependence. An example of this task is presented in Subsection~\ref{Filtering of autoregressive stochastic volatilities}.
\end{description}

\subsubsection{The reservoir computing setup and its capacity}

The reservoir computing construction that we consider in this work is based on the choice of a nonautonomous discrete-time dynamical system of the form:
 \begin{equation}
 \label{vector discretized reservoir main}
\mathbf{x}(t) = F( \mathbf{x} (t-1), \mathbf{I} (t), \boldsymbol{ \theta }) ,\quad \mbox{with} \quad t \in \mathbb{Z},\, \mathbf{x}(t),\, \mathbf{I} (t) \in \mathbb{R}^N,\, \mbox{ and }\boldsymbol{ \theta } \in \mathbb{R}^K.
  \end{equation}
The map $F: \mathbb{R}^N \times \mathbb{R}^N \times \mathbb{R}^K \longrightarrow \mathbb{R} ^N $ is called the {\bf reservoir map}. The vector $\mathbf{x}(t) $ is referred to as the {\bf neuron layer} at time $t$ and each of its components $x _i(t) $ are its {\bf neuron values}. The vector $\boldsymbol{ \theta } \in \mathbb{R}^K$ contains the set of parameters that the reservoir map depends on. The vector $\mathbf{I} (t) \in \mathbb{R}^N$ is the \textbf{input forcing} of the reservoir that is constructed out of the {\bf input signal} $\left\{z(t)\right\}_{t\in \mathbb{Z}}$, $z(t)\in \mathbb{R}$, by using an {\bf input mask} ${\bf c}  \in \mathbb{R}^N$ and by setting $\mathbf{I} (t):={\bf c}z(t)$.

\begin{example}
\label{TDR example}
\normalfont
{\bf Time-delay reservoirs (TDRs)}: This RC setup~\cite{Rodan2011, gutierrez2012} is based on the sampling of the solutions of a time-delay differential equation (TDDE) of the form
\begin{equation}
\label{time delay equation}
 \dot{x} (t)= -x(t) + f(x (t- \tau), I(t), \boldsymbol{\theta}),
\end{equation}
with {\bf  time-delay} $\tau $ and where $f : \mathbb{R}\times \mathbb{R}\times \mathbb{R}^K \rightarrow \mathbb{R} $ is called the {\bf kernel map}.
A reservoir map $F: \mathbb{R} ^N \times\mathbb{R}  ^N \times\mathbb{R}  ^K \longrightarrow \mathbb{R} ^N$ like in~\eqref{vector discretized reservoir main} can be constructed in this case by considering the Euler time-discretization of~\eqref{time delay equation} with integration step $d:= \tau/N $, with $N$ the number of neurons of the desired RC setup, namely,
\begin{equation}
\label{euler discretization}
\frac{x (t)- x (t- d)}{d}=- x (t)+ f(x (t- \tau), I(t), \boldsymbol{\theta}).
\end{equation}
Next, we use an input mask ${\bf c}\in  \mathbb{R}^N $ to multiplex the input signal over the delay period by setting $\mathbf{I}(t):={\bf c} z (t) \in \mathbb{R}^N$.
We  then organize it, as well as the solutions of~\eqref{euler discretization}, in neuron layers $\mathbf{x} (t)  $ parametrized by a discretized  time $t\in \Bbb Z $ which yields
\begin{equation}
\label{recursion euler}
x _i(t):= e^{-\xi}x_{i-1}(t)+(1-e^{-\xi})f(x _i(t-1), I _i(t), \boldsymbol{\theta}),\enspace \mbox{with} \enspace x _0(t):= x _N(t-1), \enspace \mbox{and} \enspace \xi:=\log (1+ d),
\end{equation}
where $x_i (t) $ and $I_i (t) $ stand for the $i $th-components of the vectors $\mathbf{x} (t)  $ and $\mathbf{I} (t)  $, respectively. The value $d $ is referred to as the {\bf  separation between neurons}. The reservoir map~\eqref{vector discretized reservoir main} is obtained by using~\eqref{recursion euler} in order to write down the neuron values of the layer for time $t$ in terms of those for time $t-1 $ and the current input signal value. More specifically:
\begin{equation}
\label{discretized reservoir}
\left\{
\begin{array}{rcl}
x_1 (t) &=& e  ^{- \xi } x _N (t-1) + (1- e ^{- \xi }) f(x _1 (t-1), I _1 (t), \boldsymbol{ \theta }),\\
x_2 (t) &=& e  ^{-2 \xi } x _N (t-1) + (1- e ^{- \xi }) \left\{ e  ^{- \xi }f(x _1 (t-1), I _1 (t), \boldsymbol{ \theta }) + f(x _2 (t-1), I _2 (t), \boldsymbol{ \theta })\right\},\\
  &\vdots& \\
x_N (t) &=& e  ^{-N \xi } x _N (t-1) + (1- e ^{- \xi })  \sum^{N-1}_{j = 0} e  ^{- j\xi }f(x _{N-j} (t-1), I _{N-j} (t), \boldsymbol{ \theta }),
\end{array}
\right.
\end{equation}
which corresponds to a description of the form
\begin{equation}
\label{vector discretized reservoir}
\mathbf{x}(t) = F( \mathbf{x} (t-1), \mathbf{I} (t), \boldsymbol{ \theta }) , 
\end{equation}
that uniquely determines the reservoir map $F: \mathbb{R} ^N \times\mathbb{R}  ^N \times\mathbb{R}  ^K \longrightarrow \mathbb{R} ^N$.
Physical implementations of this scheme carried out with dedicated hardware are already available and have shown excellent performances in the processing of empirical data:  spoken digit recognition~\cite{jaeger2, Appeltant2011, Larger2012, Paquot2012, photonicReservoir2013}, the NARMA model identification task~\cite{Atiya2000, Rodan2011}, continuation of chaotic time series, and volatility forecasting~\cite{GHLO2012}. 
\quad $\blacksquare$
\end{example}

As we explained in the previous subsection, a task is assigned to the RC by fixing a teaching signal $\left\{y(t)\right\}_{t\in \mathbb{Z}}$ and by minimizing the mean square error committed at the time of reproducing it with an affine combination of the reservoir output $\mathbf{x}(t)$ of the form ${\mathbf W}^\top\mathbf{x}(t)+a$, with $a\in \mathbb{R}$ and ${\mathbf W}\in \mathbb{R}^N$. The optimal pair $\left({\mathbf W}_{\rm out},a_{\rm out} \right) $ is referred to as the {\bf readout layer} and is obtained by solving the ridge (or Tikhonov) regularized regression problem 
\begin{equation}
\left({\mathbf W}_{\rm out},a_{\rm out} \right):=\mathop{\rm arg\, min}_{{\mathbf W}\in \mathbb{R}^N, a\in \mathbb{R}} \left( {\rm E}\left[ \left({\mathbf W}^\top\mathbf{x}(t)+a-y(t)  \right)^2    \right]+\lambda \left\|{\mathbf W}\right\|^2\right), \quad \lambda \in \mathbb{R},
\end{equation}
whose solution is given by
\begin{align}
\label{linear_cov_system_gamma}
 {\mathbf W}_{\rm out} = & (\Gamma(0) + \lambda \mathbb{I} _N ) ^{-1}{\rm Cov} (y (t) , \mathbf{x} (t)),\\
\label{linear_cov_system_c}
a_{\rm out} = & {\rm E}\left[ y (t) \right] - {\mathbf W}_{\rm out} ^\top \boldsymbol{\mu} _x.  
\end{align}
In this expression, $\boldsymbol{\mu} _x:= {\rm E}\left[\mathbf{x} (t)\right] $ is the expectation of the reservoir output and  
\[\Gamma(0):= {\rm E}\left[\left(\mathbf{x}(t)- \boldsymbol{\mu} _x\right)\left(\mathbf{x}(t)- \boldsymbol{\mu} _x\right) ^\top\right] \] 
is the lag-zero auto covariance exhibited by the reservoir output. We show in the next subsection that if the input signal is strictly stationary then the two moments $\boldsymbol{\mu} _x $ and $\Gamma(0) $ are time-independent.
The mean square error committed by the reservoir when using the optimal readout is:
\begin{align}
\label{error rc optimal}
&{\rm E}\left[\left( {\mathbf W}_{{\rm out}} ^{\top} \cdot {\bf x}(t)  + a_{{\rm out}} -{ y}(t)\right)^2\right]= {\mathbf W}_{{\rm out}}^\top \Gamma (0)  {\mathbf W}_{{\rm out}}+{\rm var} \left(y (t)\right) - 2{\mathbf W}_{{\rm out}}^\top  {\rm Cov} (y (t) , \mathbf{x} (t))  \nonumber \\
	&={\rm var} \left(y (t)\right)-{\rm Cov} (y (t) , \mathbf{x} (t))^\top(\Gamma(0) + \lambda \mathbb{I} _N ) ^{-1}(\Gamma (0) + 2\lambda \mathbb{I} _N) (\Gamma(0) + \lambda \mathbb{I} _N ) ^{-1}{\rm Cov} (y (t) , \mathbf{x} (t)).
\end{align}
The {\bf reservoir capacity} $C( \boldsymbol{\theta}, {\bf c}, \lambda) $ is defined as one minus the mean square error that we just computed, normalized with the variance of the teaching signal
\begin{equation}
\label{capacity formula general}
C( \boldsymbol{\theta}, {\bf c}, \lambda):=\frac{{\rm Cov} (y (t) , \mathbf{x} (t))^\top(\Gamma(0) + \lambda \mathbb{I} _N ) ^{-1}(\Gamma (0) + 2\lambda \mathbb{I} _N) (\Gamma(0) + \lambda \mathbb{I} _N ) ^{-1}{\rm Cov} (y (t) , \mathbf{x} (t))}{{\rm var} \left(y (t)\right)}.
\end{equation}
We emphasize that this capacity is a natural generalization to the context of non-independent stationary input signals of the notion introduced in~\cite{Jaeger:2002, White2004, Ganguli2008, Hermans2010, dambre2012}. Additionally, we point out that $C( \boldsymbol{\theta}, {\bf c}, \lambda)$ depends on, apart from the reservoir parameters $ \boldsymbol{\theta} $, the input mask ${\bf c} $, and the regularization constant $\lambda$, also on the task determined by the teaching signal $\left\{y(t)\right\}_{t\in \mathbb{Z}}$. It is worth noting that since the normalized error coming from~\eqref{error rc optimal} is bounded between zero and one, it is clear that
$
0 \leq C( \boldsymbol{\theta}, {\bf c}, \lambda)\leq 1.
$

\subsection{The reservoir model}
The capacities~\eqref{capacity formula general} for a reservoir of the form~\eqref{vector discretized reservoir main} are in general very difficult to compute analytically. The standard approach to determine them consists hence in fixing a triple $( \boldsymbol{\theta}, {\bf c}, \lambda)$ and in estimating the corresponding reservoir capacity $C( \boldsymbol{\theta}, {\bf c}, \lambda)$  via Monte Carlo simulations. This strategy makes the finding of the optimal parameters for a given task computationally very expensive.   

In~\cite{GHLO2014_capacity} we introduced an approximate model for the time-delay reservoirs (TDRs) presented in Example~\ref{TDR example} that made possible an analytic estimation of their capacities under a very strong independence hypothesis in the input signal; this condition was already present in the original definitions of this notion~\cite{Jaeger:2002, White2004, Ganguli2008, Hermans2010, dambre2012}. These developments provided a functional link between the TDR parameters and its performance with respect to a given reconstruction task which was used to accurately determine the optimal reservoir architecture by solving a well posed optimization problem.

In this section we extend that construction to more general RCs driven by strictly stationary input signals (see Definition~\ref{def strictly stationary}). 
The approximate model of the RC in~\eqref{vector discretized reservoir main} is obtained, as in~\cite{GHLO2014_capacity}, by partially linearizing $F$ with respect to the self-delay at a stable fixed point  $\mathbf{x} _0 \in \mathbb{R}^N$ of the autonomous system associated to \eqref{vector discretized reservoir main}. This condition means that the point $\mathbf{x} _0 \in \mathbb{R}^N$ is chosen so that $F(\mathbf{x} _0, {\bf 0}_N , \boldsymbol{\theta} ) =\mathbf{x} _0 $,  $\boldsymbol{\theta} \in \mathbb{R} ^K$, and for which the spectral radius $\rho\left( A(\mathbf{x} _0,  \boldsymbol{\theta})\right)<1 $, with 
$ A(\mathbf{x} _0,  \boldsymbol{\theta} ) :=D_{\mathbf{x} } F(\mathbf{x} _0,{\bf 0} _N, \boldsymbol{ \theta })$, in order to ensure stability. In~\cite{GHLO2014_capacity} we provided both theoretical and empirical evidence that suggests that optimal reservoir performance can be achieved when working in a statistically stationary regime around a stable equilibrium. The stability of the point $\mathbf{x} _0 $ implies, in passing, that the reservoir states $\mathbf{x} (t) $ remain close to $\mathbf{x} _0 $, and hence justifies approximating the reservoir~\eqref{vector discretized reservoir main} by the series expansion:
\begin{align}
\label{linearization first}
\mathbf{x} (t) & = F(\mathbf{x} _0, {\bf 0}_N , \boldsymbol{\theta} ) +D_{\mathbf{x} } F(\mathbf{x} _0,{\bf 0} _N, \boldsymbol{ \theta }) (\mathbf{x} (t-1) - \mathbf{x} _0)  +
\sum_{i=1}^R \dfrac{1}{i !}D_{\mathbf{I} }^{(i)} F(\mathbf{x} _0,{\bf 0} _N, \boldsymbol{ \theta })\overbrace{\mathbf{I}(t)\otimes\cdots\otimes\mathbf{I}(t)}^{i\textrm{ factors}}  \nonumber\\
&=\mathbf{x} _0+A(\mathbf{x} _0,  \boldsymbol{\theta} ) (\mathbf{x} (t-1) - \mathbf{x} _0)  + \boldsymbol{\varepsilon}  (t),
\end{align}
where $\boldsymbol{\varepsilon}  (t) \in \mathbb{R}^N$ is a vector whose entries are polynomial functions of the input signal $z(t)$ at $t$. More specifically  
\begin{align}
  \boldsymbol{\varepsilon}  (t)&=\sum_{i=1}^R \dfrac{1}{i !}D_{\mathbf{I} }^{(i)} F(\mathbf{x} _0,{\bf 0} _N, \boldsymbol{ \theta })\mathbf{I}(t)\otimes\cdots\otimes\mathbf{I}(t)=\sum_{i=1}^R \dfrac{1}{i !}D_{\mathbf{I} }^{(i)} F(\mathbf{x} _0,{\bf 0} _N, \boldsymbol{ \theta }) \left({\bf c}z(t)\right)\otimes\cdots\otimes \left({\bf c}z(t)\right)  \nonumber\\
  &=\sum_{i=1}^R \dfrac{z (t)^i}{i !}D_{\mathbf{I} }^{(i)} F(\mathbf{x} _0,{\bf 0} _N, \boldsymbol{ \theta })\underbrace{\left({\bf c}\otimes\cdots\otimes {\bf c} \right)}_{i \textrm{ factors}} 
  =\left(
  q_R^1\left(z(t),{\bf c} \right), \ldots,
  q_R^N\left(z(t),{\bf c} \right)
  \right)^\top,
  \label{definition epsilon}
  \end{align}
where $ q_R^j\left(\cdot,{\bf c}\right)$ is a polynomial of degree $R\in \mathbb{N}$ whose monomial of order $i$ has as coefficient the value 
$\dfrac{1}{i !}D_{\mathbf{I} }^{(i)} F_j(\mathbf{x} _0,{\bf 0} _N, \boldsymbol{ \theta })\underbrace{\left({\bf c}\otimes\cdots\otimes {\bf c} \right)}_{i \textrm{ factors}}$ with $F_j$ is the $j$-th component of the map $F:=\left(F_1,\dots,F_N\right)$ in \eqref{vector discretized reservoir main}. In what follows, we will refer to the recursion 
\begin{equation}
\label{linearization}
\mathbf{x} (t) =\mathbf{x} _0+A(\mathbf{x} _0,  \boldsymbol{\theta} ) (\mathbf{x} (t-1) - \mathbf{x} _0)  + \boldsymbol{\varepsilon}  (t)
\end{equation}
as the {\bf approximate reservoir model} or just the {\bf reservoir model}. We now notice that the strict stationarity of $\{z(t)\}_{t \in \Bbb Z}$ implies that of $\{\boldsymbol{\varepsilon}  (t)\}_{t \in \Bbb Z}$. In particular,
\begin{equation}
\label{mean of epsilon}
\boldsymbol{\mu}_{\boldsymbol{\varepsilon}}:= {\rm E} \left[\boldsymbol{\varepsilon}(t)\right]=
  \left(\left(q_R^1\left(x,{\bf c} \right)\right)\left(\mu^{\cdot}_z \right), \cdots,
  \left(q_R^N\left(x,{\bf c} \right)\right)\left(\mu^{\cdot}_z \right)\right)^\top,
\end{equation}
where the symbol $\left(q_R^i\left(x,{\bf c} \right)\right)\left(\mu^{\cdot}_z \right)$ stands for the evaluation of the polynomial $q_R^N\left(x,{\bf c} \right)$ according to the following convention: any monomial of the form $a_rx^r$ is replaced by $a_r\mu_z^r$. A similar convention can be used to write down the autocovariance $\Gamma_{\boldsymbol{\varepsilon}}(h)$, $h \in \mathbb{Z}$  of $\{\boldsymbol{\varepsilon}(t)\}_{t \in \Bbb Z}$. Indeed, for any $i,j \in \{1,\dots,N\}$:
\begin{equation}
\label{autocovariance of epsilon}
\left(\Gamma_{\boldsymbol{\varepsilon}}(h)\right)_{i,j}={\rm E} \left[\boldsymbol{\varepsilon}^i(t) \boldsymbol{\varepsilon}^j(t+h) \right]-\boldsymbol{\mu}_{\boldsymbol{\varepsilon}}^i\boldsymbol{\mu}_{\boldsymbol{\varepsilon}}^j =\left( q_R^i\left(x,{\bf c}\right)\bullet q_R^j\left(y,{\bf c}\right) \right) \left( \mu^{\cdot,\cdot}_z(h)\right)-\boldsymbol{\mu}_{\boldsymbol{\varepsilon}}^i \boldsymbol{\mu}_{\boldsymbol{\varepsilon}}^j,
\end{equation}
where the symbol $\bullet $ denotes polynomial multiplication and the first summand stands for the evaluation of the bivariate polynomial $ q_R^i\left(x,{\bf c}\right)\bullet q_R^j\left(y,{\bf c}\right)$ according to the following convention: any monomial of the form $a_{r,s}x^ry^s$ is replaced by $a_{r,s}\mu^{r,s}_{z}(h)$, with $\mu^{r,s}_{z} $ the second order automoment of $\{z(t)\}_{t \in \Bbb Z}$.

The following proposition, whose proof can be found in the appendix, shows that the strict stationarity of the input signal implies the second order stationarity of the output $\{ \mathbf{x}(t)\} _{t \in \Bbb Z}$ of the approximate reservoir~\eqref{linearization}.

\begin{proposition} 
\label{stationarity of approximate reservoir}
Let $\left\{\mathbf{x}(t)\right\}_{t \in \Bbb Z}$ be the output   of the reservoir model~\eqref{linearization}. Suppose that the spectral radius $\rho\left( A(\mathbf{x} _0,  \boldsymbol{\theta})\right)<1 $ and that the input signal $\{z(t)\}_{t\in \mathbb{Z}}$ is strictly stationary and has finite automoments up to order $2R $ ($R $ is the order of the expansion that defines the reservoir model~\eqref{linearization}). Under those hypotheses, the reservoir output  $\left\{\mathbf{x}(t)\right\}_{t\in \mathbb{Z}}$ is second order stationary (see Remark~\ref{second order extension}) and the corresponding time-independent mean $\boldsymbol{\mu}_{\mathbf{x}} $ and autocovariances $\Gamma $ are given by:
\begin{equation}
\label{mean of x}
\boldsymbol{\mu}_{\mathbf{x}}:= {\rm E}\left[\mathbf{x}(t)\right] =\mathbf{x}_0+\left( {\mathbb{I}  _N} - A(\mathbf{x} _0,  \boldsymbol{\theta})\right) ^{-1}\boldsymbol{\mu}_{\boldsymbol{\varepsilon}},
\end{equation}
\begin{equation}
\label{autocovariance of x}
\Gamma(h):= {\rm E}\left[   \left(\mathbf{x}(t)-\boldsymbol{\mu}_{\mathbf{x}}\right)  \left(\mathbf{x}(t+h)-\boldsymbol{\mu}_{\mathbf{x}}\right)^\top \right]=\sum_{j,k=0}^\infty A^j\Gamma_{\boldsymbol{\varepsilon}}\left(k-j-h\right)\left(A^k\right)^\top, \quad h \in \Bbb Z,
\end{equation}
with $\boldsymbol{\mu}_{\boldsymbol{\varepsilon}}$ and $\Gamma_{\boldsymbol{\varepsilon}}$ provided by \eqref{mean of epsilon} and \eqref{autocovariance of epsilon}, respectively.
Under these hypotheses, the recursion \eqref{linearization} that determines the reservoir model can be rewritten as 
\begin{equation}
\label{reservoir and stationarity}
 \left( \mathbf{x}(t)-\boldsymbol{\mu}_{\mathbf{x}} \right) =A(\mathbf{x} _0,  \boldsymbol{\theta} ) (\mathbf{x} (t-1) -\boldsymbol{\mu}_{\mathbf{x}})  + \left(\boldsymbol{\varepsilon}  (t)-\boldsymbol{\mu}_{\boldsymbol{\varepsilon}}\right),
\end{equation}
and
\begin{equation}
\label{unique solution of reservoir and stationarity}
  \mathbf{x}(t) = \boldsymbol{\mu}_{\mathbf{x}} + \sum_{j=0}^\infty A(\mathbf{x} _0,  \boldsymbol{\theta} )^j \left(\boldsymbol{\varepsilon}  (t-j)-\boldsymbol{\mu}_{\boldsymbol{\varepsilon}}\right),
\end{equation} 
is the unique stationary solution of \eqref{reservoir and stationarity}.
\end{proposition}

\subsection{Reservoir capacity estimations for signal forecasting, reconstruction, and filtering}
\label{Reservoir capacity estimations for signal forecasting, reconstruction, and filtering}

We now use the reservoir model introduced in the previous section in order to provide estimates for the different information processing tasks that we {described} in Section~\ref{The tasks}. All along this section, we assume that the input signal $\{z (t)\}_{t \in \Bbb Z}$ is strictly stationary and has finite automoments up to order $2R $ so that we can use the results contained in Proposition~\ref{stationarity of approximate reservoir}.

\subsubsection{Linear and quadratic forecasting and reconstruction} \label{subsubsec:linear and quad case}
\paragraph{The linear case.}
Consider the linear forecasting/reconstruction task $H: \mathbb{R}^{f+h+1}\rightarrow \mathbb{R}$ determined by the assignment 
\[{\bf z}^{f,h}(t)=\left(z(t+f),\dots,z(t),\dots,z(t-h)\right) \in \mathbb{R}^{f+h+1} \longmapsto \mathbf{L}^\top {\bf z}^{f,h}(t)
\]
with $\mathbf{L} \in \mathbb{R}^{f+h+1}$. We construct the teaching signal by setting $y(t):=\mathbf{L}^\top {\bf z}^{f,h}(t)$. Notice first that 
$\mu_y : = {\rm E } \left[ y(t)\right]= \mu_z \mathbf{L}^\top \mathbf{i}_{f+h+1}$. We now estimate the memory capacity $C_H( \boldsymbol{\theta}, {\bf c}, \lambda) $ associated to the task $H$ and  exhibited by the reservoir model~\eqref{reservoir and stationarity}. Notice that the evaluation of the expression~\eqref{capacity formula general} requires the computation of the lag-zero autocovariance $\Gamma (0) $ of the reservoir output in terms of the reservoir parameters, as well as ${\rm var} \left(y (t)\right) $ and ${\rm Cov} (y (t) , \mathbf{x} (t)) $. The expression for $\Gamma(0) $ is explicitly provided by~\eqref{autocovariance of x}; regarding ${\rm var} \left(y (t)\right) $ and ${\rm Cov} (y (t) , \mathbf{x} (t)) $ we have:
\begin{itemize}
\item ${\rm var} ({y} (t) ) = {\rm E } \left[ y(t)^2\right]-\mu_y^2 =\mathbf{L}^\top  {\rm E } \left[{\bf z}^{f,h}(t) {\bf z}^{f,h}(t)^\top \right]\mathbf{L}-\mu_y^2= \mathbf{L}^\top  (\Gamma^z -  \mu_z ^2 \mathbf{i}_{f+h+1}\mathbf{i}_{f+h+1}^\top)\mathbf{L}$,
with $\Gamma^z \in \mathbb{S}_{f+h+1}$ defined by 
\begin{equation}
\Gamma^z_{i,j}=\mu_{z}^{1,1}(i-j), \quad \mbox{with} \quad i,j \in \{1, \ldots, f+h+1\},
\end{equation}
and $\mu_{z}^{1,1} $ the second order automoment of the input signal.
\item ${\rm Cov}(y(t),\mathbf{x} (t))$: consider the representation in \eqref{unique solution of reservoir and stationarity} of the unique stationary solution of the reservoir model, that is, 
\begin{equation}
\label{representation used in linear case}
\left( \mathbf{x} (t)-\boldsymbol{\mu}_{\mathbf{x}}\right) = \sum_{j=0}^\infty A(\mathbf{x} _0,  \boldsymbol{\theta} )^j \boldsymbol{\rho} (t-j) ,
\end{equation}
with $\boldsymbol{\rho}(t):=\boldsymbol{\varepsilon}  (t)-\boldsymbol{\mu}_{\boldsymbol{\varepsilon} }$. Consequently 
\begin{align}
{\rm Cov}(y(t),\mathbf{x} (t))&=\sum_{j=1}^{f+h+1}{\rm Cov} \left(L_j z(t+f+1-j),\mathbf{x} (t)\right) \nonumber\\
&=\sum_{j=1}^{f+h+1} \sum_{k=0}^\infty L_j A(\mathbf{x} _0,  \boldsymbol{\theta} )^k {\rm E } \left[ \left(z(t+f+1-j)-\mu_z\right)\left(\boldsymbol{\varepsilon}(t-k)-\boldsymbol{\mu}_{\boldsymbol{\varepsilon} }\right)\right] \nonumber\\
&=\sum_{j=1}^{f+h+1} \sum_{k=0}^\infty L_j A(\mathbf{x} _0,  \boldsymbol{\theta} )^k \left[ \begin{pmatrix} \left(x\bullet q_{R}^1\left(y,{\bf c}\right)\right)\left(\mu_z^{1,\cdot}\left(f+k+1-j\right)\right)\\ \vdots \\ \left(x\bullet q_{R}^N\left(y,{\bf c}\right)\right)\left(\mu_z^{1,\cdot}\left(f+k+1-j\right)\right)  \end{pmatrix}-\mu_z\boldsymbol{\mu}_{\boldsymbol{\varepsilon} }\right] ,
\end{align} 
where this expression has been written  using the same convention as in~\eqref{autocovariance of epsilon}.
\end{itemize}
\paragraph{The quadratic case.}
Consider the quadratic forecasting/reconstruction task $H: \mathbb{R}^{f+h+1}\rightarrow \mathbb{R}$ defined by the assignment
$
{\bf z}^{f,h}(t)\longrightarrow{\bf z}^{f,h}(t)^\top Q {\bf z}^{f,h}(t),$ with $Q\in \mathbb{S}_{f+h+1}$.
We then define  the teaching signal
\begin{equation}
y(t):=H\left({\bf z}^{f,h}(t)\right)=\sum_{i,j=1}^{f+h+1} Q_{i,j}z(t+f+1-i)z(t+f+1-j).
\end{equation}
This implies that 
\begin{equation}
\label{mean of y quadratic case}
\mu_y:={\rm E } \left[ y(t)\right]= \sum_{i,j=1}^{f+h+1} Q_{i,j}\mu_{z}^{1,1}(i-j).
\end{equation}
At the same time
\begin{equation}
y(t)^2 =\sum_{i,j,k,l=1}^{f+h+1}Q_{i,j}Q_{k,l}z(t+f+1-i)z(t+f+1-j)z(t+f+1-k)z(t+f+1-l),
\end{equation}
and hence
\begin{equation}
\label{variance of y quadratic case}
{\rm E } \left[ y(t)^2\right] = \sum_{i,j,k,l=1}^{f+h+1}Q_{i,j}Q_{k,l}\mu_{z}^{1,1,1,1}(i-j,i-k,i-l).
\end{equation}
Consequently, by \eqref{mean of y quadratic case} and \eqref{variance of y quadratic case},
\begin{equation*}
{\rm var} \left(y (t)\right)= \sum_{i,j,k,l=1}^{f+h+1}Q_{i,j}Q_{k,l}\mu_{z}^{1,1,1,1}(i-j,i-k,i-l)- \left(\sum_{i,j=1}^{f+h+1}Q_{i,j}\mu_{z}^{1,1}(i-j)\right)^2 .
\end{equation*}

In order to compute $ {\rm Cov} \left( y(t),\mathbf{x} (t)\right)$, we use again the representation \eqref{representation used in linear case} and hence

\begin{align}
&{\rm Cov} \left( y(t),\mathbf{x} (t)\right) = {\rm Cov} \left( y(t),\mathbf{x} (t)-\boldsymbol{\mu}_{\mathbf{x}}\right)= \sum_{k=0}^\infty A(\mathbf{x} _0,  \boldsymbol{\theta} )^k {\rm Cov} \left( y(t),\boldsymbol{\rho} (t-k)\right) \nonumber \\
&= \sum_{k=0}^\infty A(\mathbf{x} _0,  \boldsymbol{\theta} )^k\left[ {\rm E } \left[ y(t)\boldsymbol{\varepsilon}(t-k)\right] -\mu_y \boldsymbol{\mu}_{\boldsymbol{\varepsilon}}  \right] \nonumber \\
&= \sum_{k=0}^\infty \sum_{i,j=1}^{f+h+1} A(\mathbf{x} _0,  \boldsymbol{\theta} )^k Q_{i,j}{\rm E } \left[  z(t+f+1-i)z(t+f+1-j)\boldsymbol{\varepsilon}(t-k)  \right] -\mu_y\sum_{k=0}^\infty A(\mathbf{x} _0,  \boldsymbol{\theta} )^k\boldsymbol{\mu}_{\boldsymbol{\varepsilon}}\nonumber \\
&= \sum_{k=0}^\infty \sum_{i,j=1}^{f+h+1} Q_{i,j} A(\mathbf{x} _0,  \boldsymbol{\theta} )^k \begin{pmatrix}  \left(x\bullet y\bullet q_{R}^1\left(z,{\bf c}\right)\right) \left(\mu_{z}^{1,1,\cdot}(i-j,i-k-f-1)\right)  \\ \vdots \\ \left(x\bullet y\bullet q_{R}^N\left(z,{\bf c}\right)\right) \left(\mu_{z}^{1,1,\cdot}(i-j,i-k-f-1)\right)    \end{pmatrix}-\mu_y \sum_{k=0}^\infty  A(\mathbf{x} _0,  \boldsymbol{\theta} )^k\boldsymbol{\mu}_{\boldsymbol{\varepsilon}}.
\end{align}

\subsubsection{Filtering of stochastic costationary signals}
\label{Filtration of stochastic costationary signals}

This case is a generalization of the previous one in which the input and teaching signal exhibit statistical dependence, even though they do not necessarily have a deterministic functional link. This statistical relation is used by the RC in order to construct a nonparametric estimation of the conditional expectation ${\rm E} \left[y(t)\left|\mathcal{F}_t \right.\right]$, where $\mathcal{F}_t$ is the information set generated by the input signal up to time $t$, that is, $\mathcal{F}_t =\sigma\left(z(t),z(t-1),\dots\right)$. As we already explained in Subsection~\ref{The tasks}, this conditional expectation minimizes the mean square error committed by the RC at the time of reproducing the teaching signal.
We start by introducing the following definition.

\begin{definition}
Let  $\{z(t)\}_{t \in \Bbb Z}$ and $\{y(t)\}_{t \in \Bbb Z}$ be two one-dimensional stochastic time series. Given $r \in \mathbb{N}$ and $h \in \mathbb{Z}$ we define the \textbf{higher order comoment} as 
\begin{equation}
\mu_{y,z}^r(t,h):= {\rm E}\left[ y(t)z(t+h)^r\right].
\end{equation}
If the higher-order comoments up to order $r$ exist and are time-independent, we say that $\{y(t)\}_{t \in \Bbb Z}$ and $\{z(t)\}_{t \in \Bbb Z}$ are \textbf{ $r$th-order costationary} and we note 
\begin{equation}
\mu_{y,z}^r(h):= {\rm E}\left[ y(t)z(t+h)^r\right], \quad \textrm{for any } t \in \mathbb{Z}.
\end{equation}
\end{definition}

Suppose now that $\{z(t)\}_{t\in \mathbb{Z}}$ is the input of the RC and $\{y(t)\}_{t\in \mathbb{Z}}$ is a teaching signal defining a specific filtering task. As we did all along this section, we assume that the input signal is strictly stationary and has finite automoments up to order $2R $; additionally we suppose that $\{z(t)\}_{t\in \mathbb{Z}}$ and $\{y(t)\}_{t\in \mathbb{Z}}$ are costationary of order $R $.

With these assumptions, we can explicitly spell out the performance of the RC in the filtering task by using \eqref{capacity formula general}  and by noting, first, that ${\rm var}(y(t))$ can be estimated out of the teaching signal and second, that by \eqref{representation used in linear case}:
\begin{align}
{\rm Cov} &\left( y(t),\mathbf{x} (t)\right)={\rm Cov} \left( y(t),\mathbf{x} (t)-\boldsymbol{\mu}_{\mathbf{x}}\right)= \sum_{j=0}^\infty A(\mathbf{x} _0,  \boldsymbol{\theta} )^j {\rm Cov} \left( y(t),\boldsymbol{\varepsilon} (t-j)-\boldsymbol{\mu}_{\boldsymbol{\varepsilon}}\right)\nonumber\\
&=\sum_{j=0}^\infty A(\mathbf{x} _0,  \boldsymbol{\theta} )^j {\rm Cov} \left( y(t),\boldsymbol{\varepsilon} (t-j)\right)=\sum_{j=0}^\infty A(\mathbf{x} _0,  \boldsymbol{\theta} )^j \left[\begin{pmatrix} \left(x\bullet q_R^1\left(u,{\bf c}\right)\right) \left(\mu_{y,z}^{\cdot}(-j)\right) \\ \vdots \\  \left(x\bullet q_R^N\left(u,{\bf c}\right)\right) \left(\mu_{y,z}^{\cdot}(-j)\right) \end{pmatrix}-\mu_z\boldsymbol{\mu}_{\boldsymbol{\varepsilon}}\right], \label{cov filtering}
\end{align}
where the expression $\left(x\bullet q_R^i\left(u,{\bf c}\right)\right) \left(\mu_{y,z}^{\cdot}(-j)\right)$ stands for the evaluation of the polynomial $x\bullet q_R^i\left(u,{\bf c}\right)$ on the variables $x$ and $u$, according to the following convention: each monomial of the form $axu^r$ is replaced by $a\mu_{y,z}^r(-j)$.

We emphasize that given the input and teaching signals $\{z(t)\}_{t\in \mathbb{Z}}$ and $\{y(t)\}_{t\in \mathbb{Z}}$, respectively, the higher order comoments can be estimated out of the training sample and inserted in \eqref{cov filtering}. When this expression, together with the estimate of  ${\rm var} \left(y (t)\right)$ is substituted in \eqref{capacity formula general}, we obtain an estimate of the RC capacity for any value of its parameters $  \boldsymbol{\theta}$ and the input mask ${\bf c}$.

 \subsection{The fading memory and the separation properties}
 
The fading memory and the separation properties that we describe later on in this section have been identified in the context of reservoir computing to be in relation with good information processing performances (see~\cite{Yildiz2012, lukosevicius} and references therein). The goal of the following paragraphs is showing that the reservoir model~\eqref{linearization} for the discrete-time reservoir computer~\eqref{vector discretized reservoir main} exhibits these features under reasonable assumptions on the reservoir map.
 \begin{definition}
 Consider the discrete-time reservoir map \eqref{vector discretized reservoir main}. 
 \begin{description}
 \item[{\bf (i)}] We say that the reservoir map \eqref{vector discretized reservoir main} satisfies the {\bfi  uniform fading memory property (UFMP)} whenever for any $\varepsilon > 0$ there exist $\delta_{\varepsilon } >0$ and $h_{\varepsilon } \in \mathbb{N} $ such that if for any two input signals $\left\{ z (t)\right\}_{t\in \Bbb Z } $, $\left\{ z' (t)\right\}_{t\in \Bbb Z } $ the relation $|z (s)-z' (s)|< \delta_{\varepsilon } $ holds for all $s \in \left[ t-h_{\varepsilon }, t\right]  $, $t \in \Bbb Z $, then the corresponding outputs $\mathbf{x} (t)$, $\mathbf{x} ' (t)$ are such that $||\mathbf{x} (t) - \mathbf{x} ' (t) ||< \varepsilon $. The values $\delta_{\varepsilon } >0$ and $h_{\varepsilon } \in \mathbb{N} $ corresponding to a given $\varepsilon > 0 $ are the same for any $t \in \Bbb Z$.
\item  [{\bf (ii)}] We say that \eqref{vector discretized reservoir main} satisfies the {\bfi  separation  property (SP)} if for two input signals $\left\{ {z} (t) \right\} _{t\in \Bbb Z }  $, $\left\{ {z}' (t) \right\} _{t\in \Bbb Z }  $ that differ only at some time point $s \in \mathbb{Z} $, that is, $z (s) \neq z '(s)$,   the corresponding outputs satisfy that $ \mathbf{x} (t) \neq \mathbf{x} ' (t) $ for any $t \ge s$.
 \end{description} 
 \end{definition}
 
 The proof of the following two results can be found in the appendices at the end of the paper.

 \begin{theorem}
 \label{theorem sp ufm properties} 
Consider the reservoir model \eqref{linearization} driven by the real valued and non-necessarily stationary input signal $\left\{ z  (t)\right\}_{t \in \Bbb Z }$. 
 \begin{description}
 \item[{\bf (i)}]  Let ${\bf c} \in \mathbb{R} ^N  $ be an input mask and $ \mathbf{I} (t):= {\bf c} z (t) $ the corresponding input forcing. Let  
 \[
F_I^R( \mathbf{I}  (t), \mathbf{x} _0, \boldsymbol{\theta} ):= \sum_{i=1}^R \dfrac{1}{i !}D_{\mathbf{I} }^{(i)} F(\mathbf{x} _0,{\bf 0} _N, \boldsymbol{ \theta })\overbrace{\mathbf{I}(t)\otimes\cdots\otimes\mathbf{I}(t)}^{i\textrm{ factors}}
 \]
 be the $R$th-order Taylor series expansion of the reservoir map $F$ at the point $(\mathbf{x} _0,{\bf 0} _N, \boldsymbol{ \theta })$ with respect to the input forcing $\mathbf{I} (t) $. Assume that one of the following conditions holds: 
 \begin{description}
\item  [{\bf (a)}] The map $F^R _I(\cdot , \mathbf{x} _0, \boldsymbol{ \theta }): \mathbb{R} ^N\rightarrow \mathbb{R}^N $ is injective.
\item  [{\bf (b)}] The input signal is bounded, that is, there exists $k \in \mathbb{R} ^+$ such that  $|z (t) |<k$, for all $t\in \Bbb Z $, and the map $F^R _I$ is injective in the set $\mathcal{B}= \left\{ \mathbf{I} \in \mathbb{R}^N | \enspace \|\mathbf{I}\|< \|{\bf c}\| k \right\} $. 
\end{description} 
If additionally, the linear map $ A(\mathbf{x} _0,  \boldsymbol{\theta} ) :=D_{\mathbf{x} } F(\mathbf{x} _0,{\bf 0} _N, \boldsymbol{ \theta }): \mathbb{R}^N \rightarrow  \mathbb{R}^N$ has no zero eigenvalues, then the reservoir model satisfies the separation property.
 \item[{\bf (ii)}]  Suppose that the input signal $\{z(t)\}_{t\in \mathbb{Z}}$ is strictly stationary with finite automoments up to order $2R $ and that it is bounded, that is, there exists $k \in \mathbb{R} ^+$ such that  $|z (t) |<k$ for all $t\in \Bbb Z $. If additionally  the linear map $ A(\mathbf{x} _0,  \boldsymbol{\theta} ) $ is such that $\|A(\mathbf{x} _0,  \boldsymbol{\theta} )\|  <1$, with $\| \cdot \|  $ some matrix norm induced from $\mathbb{R}^N $, then the reservoir model \eqref{linearization} satisfies the uniform fading memory property.  
\end{description}

\end{theorem}

\begin{remark}
\normalfont
This result can be easily extended to multidimensional input signals, that is, $\left\{ {\bf z}  (t)\right\}_{t \in \Bbb Z }$, ${\bf z} (t) \in \mathbb{R}^n $. In that case (see~\cite{RC3}  for the details) the RC is constructed by using an input mask ${\bf c}\in \mathbb{M} _{N,n}$ that takes care not only of the temporal, but also of the dimensional multiplexing by setting $\mathbf{I}(t):= {\bf c} {\bf z} (t) $. The only additional hypothesis needed in that situation is that the rank of ${\bf c} $ has to equal $n$ in order to conclude part {\bf (ii)} of the theorem.
\end{remark}

The following result contains a statement analogous to that of Theorem~\ref{theorem sp ufm properties} in the particular case of the time-delay reservoirs   introduced in Example~\ref{TDR example}. In that situation, some hypotheses are either automatically satisfied or can be formulated in a simplified manner.

 \begin{corollary}
 \label{corollary sp ufm properties} 
 Consider a time-delay reservoir of the type introduced in Example~\ref{TDR example} with nonlinear kernel $f : \mathbb{R}\times \mathbb{R}\times \mathbb{R}^K \rightarrow \mathbb{R} $, parameters $\boldsymbol{\theta} \in \mathbb{R} ^K $, and a non-necessarily stationary input signal $\left\{ z  (t)\right\}_{t \in \Bbb Z }$, $z (t) \in \mathbb{R} $. 
 \begin{description}
 \item[{\bf (i)}] Let ${\bf c} \in \mathbb{R} ^N  $ be an input mask and $ \mathbf{I} (t):= {\bf c} z (t) $ the corresponding input forcing. Let $f_I^R( I  (t), {x} _0, \boldsymbol{\theta} ):= \sum^{R}_{i = 1} \dfrac{1}{i!} (\partial _I^{(i)} f)({x} _0 , 0, \boldsymbol{\theta} ) I  (t) ^i$ be the $R$th-order Taylor series expansion of the kernel map $f$ at the point $({x} _0 , 0, \boldsymbol{\theta})$ with respect to the input forcing $I (t) $. Assume that one of the following conditions holds: 
 \begin{description}
\item  [{\bf (a)}] The map $f^R _I(\cdot, {x} _0, \boldsymbol{\theta}  ): \mathbb{R} \rightarrow \mathbb{R} $ is injective;
\item  [{\bf (b)}] The input signal is bounded, that is, there exists $k \in \mathbb{R} ^+$ such that  $|z (t)|<k$, for all $t\in \Bbb Z $, and the map $f^R _I(\cdot, {x} _0, \boldsymbol{\theta}  )$ is injective in the set $\mathcal{B}= \left\{ I \in \mathbb{R} \mid |I|< \|C\| k \right\} $. 
\end{description} 
Then the corresponding  TDR model satisfies the {\rm {\bf (SP)}}.
 \item[{\bf (ii)}]  Suppose that the input signal $\{z(t)\}_{t\in \mathbb{Z}}$ is strictly stationary with finite automoments up to order $2R $ and that it is bounded, that is, there exists $k \in \mathbb{R} ^+$ such that  $|z (t) |<k$, for all $t\in \Bbb Z $. If the partial derivative $\partial _x f (x _0 , 0, \boldsymbol{\theta} )$ of the nonlinear kernel $f$ evaluated at the point $ (x _0 , 0, \boldsymbol{\theta} )$ satisfies the condition $|\partial _x f (x _0 , 0, \boldsymbol{\theta} )|<1$, then the  TDR model satisfies the {\rm {\bf (UFMP)}}.  
\end{description} 
\end{corollary}

\section{Examples}
\label{Examples}

This section describes examples in connection with the three different tasks listed in Subsection~\ref{The tasks}, namely, forecasting, reconstruction, and filtering. The first two examples take place in the context of two parametric stochastic time series families (ARMA and GARCH). As we will see, forecasting in those two setups comes down, in the terminology introduced in Section~\ref{Reservoir capacity estimations for signal forecasting, reconstruction, and filtering}, to solving a linear and a quadratic forecasting task for which the performance of RC has been already empirically evaluated in~\cite{GHLO2012}. The last example falls in the category of pure filtering. In that case we will show how a specific type of RC is capable of outperforming two standard filtering techniques (Kalman filtering and the hierarchical-likelihood approach). Additionally, we will evaluate the accuracy of the capacity formulas introduced in Section~\ref{Reservoir capacity estimations for signal forecasting, reconstruction, and filtering} and based on the reservoir model~\eqref{linearization} at the time of estimating the performance of the actual RC.

\subsection{Multistep forecast of ARMA temporal aggregates}
\label{Multistep forecast of ARMA temporal aggregates}

Consider the causal and invertible ARMA(p,q) specification (see~\cite{Box1976, BrocDavisYellowBook}, and references therein for details) determined by the equivalent relations 
\begin{equation}
\boldsymbol{\Phi}(L)z(t)=\boldsymbol{\Theta}(L) \zeta(t),\; z(t)=\sum_{i=0}^\infty\psi_i\zeta(t-i),\;  \zeta(t)=\sum_{j=0}^\infty \pi_{j}z(t-j), \quad \{\zeta(t)\} \sim {\rm IID} (0,\sigma^2),
\end{equation}
where $L$ is the backward shift operator defined by $L(z (t)):= z (t-1) $, $\mathbf{\Phi}(z):=1- \phi _1z - \cdots - \phi _pz ^p $, $\mathbf{\Theta}(z):=1+ \theta _1z + \cdots + \theta _qz ^q $,   $\boldsymbol{\Psi} (z):=  \mathbf{\Phi}(z) ^{-1}\mathbf{\Theta}(z) $, $\boldsymbol{\Pi} (z):=  \mathbf{\Theta}(z) ^{-1}\mathbf{\Phi}(z) $, and the symbols $\boldsymbol{\Phi}(L) $ and  $\boldsymbol{\Theta}(L) $ stand for the operators $\mathbf{\Phi}(L):=1- \phi _1L - \cdots - \phi _pL ^p $ and $\mathbf{\Theta}(L):=1+ \theta _1L + \cdots + \theta _qL ^q $, respectively.
Consider now an \textbf{aggregation vector} $\mathbf{w}\in \mathbb{R}^f$. It can be shown~\cite{GO1, GO1bis} that given a realization ${\bf z} _T $ of the process $\{z(t)\}_{t \in }$ up to time $T$, the best forecast $\widehat{z^{\mathbf{w}}}(T+f)$ (in the sense that it minimizes the mean square forecasting error) of the temporal aggregate $z^{\mathbf{w}}(T+f)= \sum_{i=1}^f w_{f-i+1} z(T+i)$ based in the information set generated by ${\bf z}_T$, is given by 
\begin{equation}
\widehat{z^{\mathbf{w}}}(T+f)= \sum_{i=1}^f \sum_{j=1}^{T+i-1+r} w_{f-i+1}\psi_j \zeta(T+i-j)=\sum_{i=1}^f \sum_{j=1}^{T+i-1+r}\sum_{k=0}^\infty w_{f-i+1}\psi_j z(T+i-j-k),
\end{equation}
with $r:={\rm max}\{p,q\}$. The mean square forecasting error ${\rm MSFE}\left( \widehat{ z^\mathbf{w}}(T+f)  \right)$ associated to this forecast is~\cite{GO1}:
\begin{eqnarray*}
\label{characteristic error 1}
{\rm MSFE}\left( \widehat{ z^\mathbf{w}}(T+f) \right)  &=& E \left[ \left(  \widehat{ z^\mathbf{w}}(T+f) -  z^\mathbf{w}(T+f)  \right) ^2 \right] \\
&=& \sigma ^2 \left[ \sum^{f}_{i = 1} w _{f-i+1} ^2  \sum^{i-1}_{l = 0} \psi _l ^2 + 2 \sum^{f-1}_{i = 1} \sum^{f}_{j = i + 1} w _{f-i+1} w _{f-j+1}  \sum^{i-1}_{l = 0} \psi _l  \psi _{ j-i+l} \right].
\end{eqnarray*}

This task can be solved via RC by using as input signal either the innovations $\zeta(t)$ or the time series values $z(t)$ up to time $T$ and using as teaching signal the values $z^{\mathbf{w}}(t)$, also up to time T. In the terminology introduced in Section~\ref{Reservoir capacity estimations for signal forecasting, reconstruction, and filtering}, in both cases this forecasting problem amounts to a linear task. For example if $z(t)$ is used as teaching signal, the linear task map $H: \mathbb{R}^f \rightarrow \mathbb{R} $ is given by 
${\bf z}^{f}(t)=\left(z(t+f),\dots,z(t+1)\right)\longmapsto  \mathbf{w}^\top{\bf z}^{f}(t) \\
$

The RC performance can be evaluated in this case by using the formula \eqref{capacity formula general} and the elements introduced in Section \ref{subsubsec:linear and quad case}, as long as the necessary automoments exist. That is the case whenever the innovations $\{\zeta(t)\}_{t \in \Bbb Z}$ are Gaussian because in that situation the input signal $\{z(t)\}_{t \in \Bbb Z}$ is a Gaussian process (see~\cite{BrocDavisYellowBook}) and hence all the automoments are finite and can be readily computed~\cite{Holmquist1988, Gaussian_moments2003}.

\subsection{Multistep forecast of temporally aggregated GARCH volatilities}
\label{Multistep forecast of temporally aggregated GARCH volatilities}

In this section we study the forecasting of the volatility associated to a flow-type aggregated sample generated by a GARCH(1,1) process~\cite{engle:arch, bollerslev:garch} and we show that it can be encoded as a quadratic forecasting task of the type described in Section~\ref{subsubsec:linear and quad case}. More specifically, consider the process $\{ z (t)\}_{t \in \Bbb Z} $ determined by  
\begin{equation}
\label{GARCH process def}
z(t)=\sigma(t)\zeta(t) \quad\textrm{with} \quad \zeta(t)\sim {\rm IID}(0,\sigma^2) \quad\textrm{and}\quad \sigma^2(t)=\alpha_0+\alpha_1 z(t-1)^2 +\beta \sigma(t-1)^2,
\end{equation}
where $\sigma(t) $ stands for the positive square root of $\sigma^2(t)$. The constants $\alpha_0,\alpha_1$ and $\beta$ are positive real numbers subjected to the constraints $\alpha_0>0$, $\alpha_1,\beta\geq0$ and $\alpha_1+\beta<1$ that ensure the existence of a unique stationary solution and the positivity of the conditional variance process $\{\sigma(t)^2\}_{t \in \Bbb Z} $ that is by construction predictable. GARCH processes are profusely used in the financial econometrics literature in the modeling of the conditional volatility associated to the time evolution of asset returns.

A problem of much importance in financial risk and portfolio management applications is the  forecasting of the variance ${\rm var}_T \left(z_{T+f}[f]\right)$ of the flow aggregate $z_{T+f}[f]:=z(T+1)+\cdots+z(T+f)$ based on the information set $\mathcal{F} _T $ generated by a realization ${\bf z}_T $ of the process $\{z(t)\}_{t \in \Bbb Z}$ up to time $T$ and the conditional volatility $\sigma(T)$.
It can be shown~\cite{henriques:ortega} that in the GARCH(1,1) context this forecast is given by 
\begin{equation}
{\rm var}_T \left(z_{T+f}[f]\right)= {\rm E}\left[z_{T+f}[f]^2\mid \mathcal{F} _T\right]\frac{f \alpha_0}{1-(\alpha_1+\beta)}+\left(\sigma(T+1)^2-\cfrac{\alpha_0}{1-(\alpha_1+\beta)}\right)\cfrac{1-(\alpha_1+\beta)^f}{1-(\alpha_1+\beta)},
\end{equation}
where $\sigma(T+1)=\left(\alpha_0+\alpha_1z(T)^2+\beta\sigma(T)^2\right)^{\frac{1}{2}}$. The estimation of this forecast can be solved via RC by using $\{z(t)\}_{t \in \Bbb Z}$ as input signal and corresponding teaching signal 
\begin{equation}
\label{teaching GARCH}
y(t):=\left(z(t+1)+ \cdots+z(t+f) \right)^2= \sum_{k_1+k_2+\cdots +k_f = 2} {2 \choose k_1, k_2, \ldots, k_f}
z(t+1)^{k_1} z(t+2)^{k_2} \cdots z(t+f)^{k_m},
\end{equation}
where the summation in the second equality is taken over all sequences of nonnegative integer indices $k  _1, \ldots, k _f$, such that the sum $k  _1+\cdots + k _f=2$. This can be encoded as a quadratic forecasting task with a task matrix $Q \in \mathbb{S}_{f}$ whose coefficients are given by the expression~\eqref{teaching GARCH}. The evaluation of the RC performance in this task using the formula \eqref{capacity formula general} and Section~\ref{subsubsec:linear and quad case} is only possible when the higher order automoments of the process~\eqref{GARCH process def} exist. This feature is by no means guaranteed in the GARCH context and imposes various semialgebric constraints on the coefficients $\alpha_0,\;\alpha_1,\; \beta$. See \cite{Li2001, Ling2002} for a characterization of the higher order moments of the GARCH family.

Time-delay reservoir (TDR) computers have shown in~\cite{GHLO2012} excellent empirical performances at the time of carrying out this task when using as data generating process the VEC-GARCH family introduced in~\cite{bollerslev:vec}, which is a multivariate generalization of the GARCH family. In that work, TDRs were also shown to outperform standard parametric multivariate volatility models in the forecasting of actual market realized volatility.

\subsection{Filtering of autoregressive stochastic volatilities}
\label{Filtering of autoregressive stochastic volatilities}

In this example we consider the autoregressive stochastic volatility (ARSV) model~\cite{Taylor:Book1} determined by the linear state-space prescription
\begin{equation}
\label{arsv model}
\left\{
\begin{array}{rcl}
z (t) &= &r+ \sigma (t) \zeta (t), \qquad \{ \zeta (t)\}_{t \in \Bbb Z} \sim {\rm IID}(0,1)\\
b (t)&= & \lambda+ \alpha b(t-1)+ w (t),  \qquad \{ w (t)\}_{t \in \Bbb Z} \sim {\rm IID}(0,\sigma _w^2)
\end{array}
\right.
\end{equation}
where $b (t):= \log (\sigma(t)^2)$, $\lambda  $ is a real parameter, and $\alpha \in (-1,1) $. {We will additionally assume that the innovations $\{ \zeta (t)\}_{t \in \Bbb Z}  $ and $\{ w (t)\}_{t \in \Bbb Z}$ are independent.} It is easy to prove that the unique stationary  process $ \{ z (t)\}_{t \in \Bbb Z}$ induced by~(\ref{arsv model}) and available in the presence of the constraint $\alpha \in (-1,1) $ is a white noise (the returns have no autocorrelation) with finite moments of arbitrary order.  Moreover, the unconditional variance $\sigma _b ^2 $ of the stationary process $\{ b (t) \} $ is given by
\begin{equation*}
\sigma _b ^2= \frac{\sigma_w ^2}{1- \alpha^2},
\end{equation*}
{and if the innovations $\{ \zeta (t)\}  $ and $\{ w (t)\}$ are Gaussian,} then the unconditional variance and kurtosis of the process $\{y (t)\} $ are given by
\begin{equation}
\label{variance and kurtosis}
{\rm var}(z (t))= {\rm E} [ \sigma (t)^2]= \exp\left[ \frac{\lambda}{1- \alpha}+ \frac{1}{2} \sigma _b ^2\right], \quad \mbox{and} \quad {\rm kurtosis}\,(z (t))= 3\exp \left( \sigma _b ^2 \right).
\end{equation}
Moreover, it can be shown~\cite{Taylor:Book1} that whenever $\sigma _b ^2 $ is small and/or $\alpha$ is close to one then the autocorrelation $\gamma (h) $  of the squared returns at lag  $h$ can be approximated by
\begin{equation*}
\gamma (h)\simeq \frac{\exp(\sigma_b ^2)-1}{3\exp(\sigma_b ^2)-1}\alpha ^h.
\end{equation*}

The main difference of the ARSV model~(\ref{arsv model}) with respect to the GARCH family is that, in this case, the volatility process $\{ \sigma (t)\}_{t \in \Bbb Z} $ is a non-observable, non-predictable stochastic latent variable that cannot be written as a function of previous realizations of the observable variable $z (t)$ and the volatilities $\sigma (t) $. Many procedures have been developed over the years to go around this difficulty whose solution is needed, in particular, to estimate the model parameters. In this section we will focus in only two them that are profusely used in the literature. First, the specific form of the  prescription~\eqref{arsv model} corresponds to a state-space model in which the observation equation is the one that yields $\{z (t)\}_{t \in \Bbb Z} $ and the state equation rules the time evolution of $b (t):= \log (\sigma(t)^2)$. This observation makes appropriate the use of the 
Kalman filter~\cite{harvey:ruiz:shephard} to obtain estimations of the conditional log-variances $b (t)  $ based on the observed values $z (t)$.
The other method that we will use as a benchmark is the hierarchical-likelihood method~\cite{lee:nelder:1, lee:nelder:2, delcastillo:lee:1, delcastillo:lee:2} (abbreviated in what follows as h-likelihood) that incorporates the unobserved volatilities as an unknown variable at the time of writing a likelihood that is optimized and that takes into account the observed time series values $z (t) $.

In the RC context, the problem of estimating the unobserved volatility $\sigma (t) $ out of the observed values of $z (t) $ up to time $t$, can be easily encoded  as a filtering problem of the type characterized in Section~\ref{The tasks} and for which the RC performance was studied in Section~\ref{Filtration of stochastic costationary signals} by using the reservoir model. Indeed, it suffices to take $z (t) $ as input signal and as teaching signal $y (t) $ the functional form of the volatility that we are interested in. Both the Kalman filter and the h-likelihood methods are designed to produced optimal (linear in the case of Kalman) estimations of the the log-variance $\log(\sigma(t)^2) $, which is a limitation to which RC is not exposed.

In the paragraphs that follow we carry out an empirical exercise in this context in order to compare the performance of the RC with that of Kalman and h-likelihood, and also to evaluate the accuracy of the capacity formulas introduced in Section~\ref{Reservoir capacity estimations for signal forecasting, reconstruction, and filtering} and based on the reservoir model~\eqref{linearization} at the time of estimating the performance of the actual RC.

We proceed by using a time-delay reservoir of the type described in the Example~\ref{TDR example} constructed with the so-called Ikeda kernel map given by the expression:
\begin{equation}
\label{Ikeda nonlinear kernel}
f(x,I, \boldsymbol{\theta})= \eta \sin ^2 \left(x+ \gamma I+ \phi\right), \enspace \boldsymbol{\theta}:=(\eta, \gamma, \phi) \in \mathbb{R} ^3.
\end{equation}
The architecture of the reservoir chosen contains 40 neurons and an input mask ${\bf c} \in \mathbb{R} ^N  $ that was randomly constructed with values uniformly distributed in the interval $[-1,1]$. 

We present to this TDR the filtering tasks consisting on estimating four different functions of the volatility $\sigma(t) $ generated by an ARSV model with parameters $r=3.9\cdot 10^{-4}$, $\sigma _w=0.675$, $\lambda= -0.821 $, and $\alpha=0.9 $. The four different teaching signals used are $y _1(t):= \sigma(t) $, $y _2(t):= \sigma(t)^2 $, $y _3(t):=\log( \sigma(t)) $, and $y _4(t):= \log(\sigma(t)^2) $.
Given a fixed input mask ${\bf c} $, the reservoir parameters $\boldsymbol{\theta} $ are optimized with respect to each of these four filtering tasks. In this case, the optimal parameters were the same for the four cases, namely, $\gamma=2.866$, $\phi=1.124$, $\eta=0.461$, and $d=0.839$; we recall that $d:= \tau/N $ is the separation between neurons. Table~\ref{performances rc volatility table} presents the performances (in terms of the normalized mean square error (NMSE)) exhibited by the TDR in the execution of the four filtering tasks and compares them with those attained using the Kalman filter and the h-likelihood approaches. The figures in the table show that these two benchmarks outperform the RC at the time of filtering the functions of the volatility (logarithm) that they have been designed for but when it comes to providing the values of the actual instantaneous volatility or variance, it is the RC that performs the best.
\begin{table}[!htp]
 \vspace{0cm}
 \noindent
\makebox[\textwidth]{%
\begin{tabularx}{1.05\textwidth}{X}
\scalebox{.78}{
\begin{tabular}{ll cccc}
\toprule
\multicolumn{6}{c}{{\bf Stochastic volatility filtering performance (NMSE)}} \\
			\midrule
	& &\multicolumn{4}{c}{{\bf Teaching signal proposed/Task solved}}\\
	\cmidrule(r){3-6}
	&& 	Instantaneous  & Instantaneous  & log of Instantaneous  & log of  Instantaneous \\
		&& 	 volatility &  variance &  volatility &  variance\\
			\midrule
	\multirow{2}{*}{{\bf Filtering Method}}& h-likelihood& 0.476& 0.730&{\bf 0.411}&{\bf 0.411}\\
	& Kalman& 0.536&0.812&0.429&0.429\\
		\midrule
	\multirow{2}{*}{{\bf Reservoir Method}}& Reservoir computer (TDR)& {\bf 0.437}& {\bf 0.594}&0.655&0.655\\
	& Reservoir model & 0.453& 0.661& 0.652& 0.601\\
\bottomrule
\end{tabular}}
\end{tabularx}}
\caption{Performances (in terms of the normalized mean square error (NMSE)) exhibited by the TDR in the execution of  four volatility filtering tasks  compared with those attained using the Kalman filter and the h-likelihood approaches.
}
\label{performances rc volatility table}
\end{table}

\begin{figure}[!ht]
\hspace*{-1.8cm}\includegraphics[scale=.40]{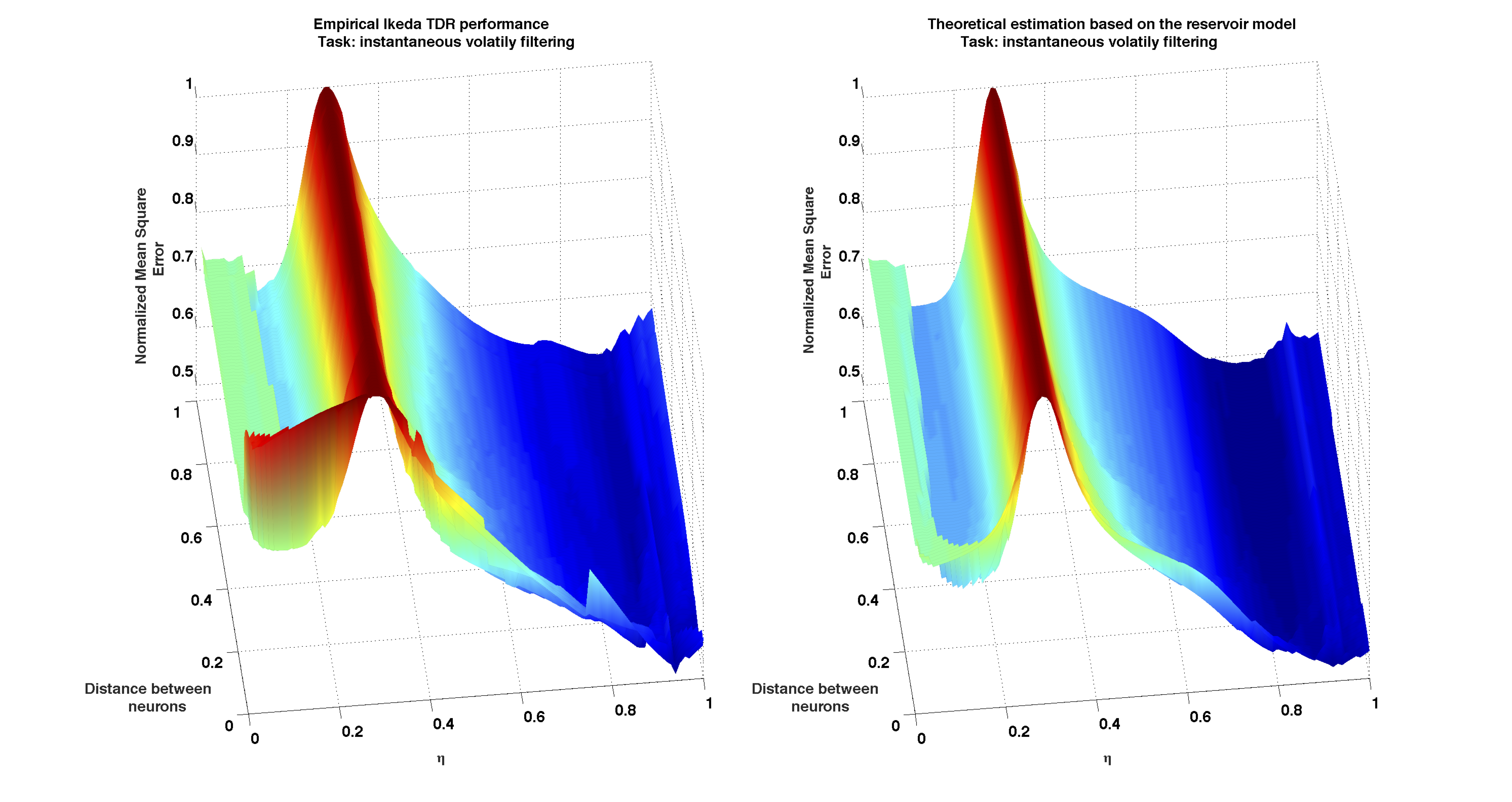}
\caption{Error surfaces associated to the filtering of the instantaneous volatility $\sigma (t) $ an ARSV data generating process. The left panel has been computed using Monte Carlo simulations in order to empirically evaluate the filtering error of the Ikeda TDR as a function of $\eta$ in~\eqref{Ikeda nonlinear kernel} and of the distance between neurons. The right panel was obtained by evaluating the formula~\eqref{capacity formula general} based on the reservoir model~\eqref{linearization} with a nonlinearity of order $R=8 $. The two surfaces have minima at virtually the same parameter values.}
\label{performances rc volatility} 
\end{figure}

We finally evaluate in the context of this filtering task the accuracy of the capacity formulas introduced in Section~\ref{Reservoir capacity estimations for signal forecasting, reconstruction, and filtering}. Figure~\ref{performances rc volatility} depicts the error surfaces associated to the filtering of the instantaneous volatility $\sigma (t) $ of the same ARSV data generating process that we considered in the construction of Table~\ref{performances rc volatility table}. The left panel has been computed using Monte Carlo simulations in order to empirically evaluate the filtering error of the Ikeda TDR as a function of the parameter $\eta$ in~\eqref{Ikeda nonlinear kernel} and of the distance between neurons. The right panel was obtained by evaluating the formula~\eqref{capacity formula general} based on the reservoir model~\eqref{linearization} with the help of the elements introduced in Section~\ref{Filtration of stochastic costationary signals} and a nonlinearity of order $R=8 $. The two surfaces clearly resemble each other and, more importantly, exhibit their minima at virtually the same parameter values. This proves that, as it was already shown in~\cite{GHLO2014_capacity, RC3} for independent signals, that  the theoretical model can be efficiently used to determine the optimal reservoir architecture in the presence of strictly stationary inputs.

\section{Appendices}
\subsection{Proof of Proposition~\ref{stationarity of approximate reservoir}}

We start by emphasizing that the potential lack of independence in the input signal $\{z (t)\}_{t \in \Bbb Z}$ implies that $\{\boldsymbol{\varepsilon}(t)\}_{t \in \Bbb Z} $ is in general not a white noise. Consequently, unlike the situation encountered in~\cite{GHLO2014_capacity, RC3}, the recursion \eqref{linearization} does not determine a standard VAR(1) model. Nevertheless, since by hypothesis $\rho\left( A(\mathbf{x} _0,  \boldsymbol{\theta})\right)<1$, the proof of the existence of a unique stationary solution for a VAR(1) model can be mimicked in this case (see, for example,  Section 2.1.1 and Proposition C.9 in~\cite{luetkepohl:book}) in order to show that the recursion
\begin{equation} 
\label{linearization not white noise}
\mathbf{x}(t)-\mathbf{x}_0  =A(\mathbf{x} _0,  \boldsymbol{\theta} ) (\mathbf{x} (t-1) -\mathbf{x}_0)  + \boldsymbol{\varepsilon}  (t),
\end{equation}
has a unique second order stationary solution given by 
\begin{equation} 
 \label{second order stationary solution}
 \mathbf{x}(t) =\mathbf{x} _0 + \sum_{i=0}^\infty A(\mathbf{x} _0,  \boldsymbol{\theta} )^i \boldsymbol{\varepsilon}  (t-i).
\end{equation}
Taking expectations in both sides of \eqref{second order stationary solution} and using the stationarity of $\boldsymbol{\varepsilon}(t)$ yields \eqref{mean of x} (see also Proposition C.10 in~\cite{luetkepohl:book}). It is straightforward  to verify using that expression that \eqref{linearization not white noise} is identical to \eqref{reservoir and stationarity} whose unique stationary solution is given by \eqref{unique solution of reservoir and stationarity} and which hence coincides necessarily with \eqref{second order stationary solution}. Finally, using the definition of $\Gamma$ and \eqref{unique solution of reservoir and stationarity}, the expression \eqref{autocovariance of x} follows. \quad $\blacksquare$

\subsection{Proof of Theorem~\ref{theorem sp ufm properties}}

\noindent {\bf Proof of part {\bf  (i)}} We start by modifying the notation introduced in~\eqref{definition epsilon} in order to specifically indicate the dependence of $ \boldsymbol{\varepsilon}  (t) $ on the input signal and on the input mask. We set:
\begin{equation}
\label{notation epsilon}
\boldsymbol{\varepsilon}(z, {\bf c}):=\sum_{i=1}^R \dfrac{z^i}{i !}D_{\mathbf{I} }^{(i)} F(\mathbf{x} _0,{\bf 0} _N, \boldsymbol{ \theta })\underbrace{\left({\bf c}\otimes\cdots\otimes {\bf c} \right)}_{i \textrm{ factors}}.
\end{equation}
Note now that  the map $\boldsymbol{\varepsilon}( \cdot , {\bf c}): \mathbb{R} \longrightarrow \mathbb{R} ^N $ that assigns $\boldsymbol{\varepsilon} (z , C)$ to the input signal $z $ is the composition of $
z \mapsto {\bf c} z $
and the map $F^R _I(\cdot , \mathbf{x} _0, \boldsymbol{ \theta }): \mathbb{R} ^N\rightarrow \mathbb{R}^N $ in the statement. Consequently, if  $ F^R _I(\cdot , \mathbf{x} _0, \boldsymbol{ \theta }) $ satisfies any of the two injectivity hypotheses in {\bf (a)} or {\bf (b)} in the statement, then the map $\boldsymbol{\varepsilon}( \cdot , {\bf c}) $ is necessarily injective, that is,
\begin{equation}
\label{injective epsilon}
\boldsymbol{\varepsilon}( z , {\bf c}) \neq \boldsymbol{\varepsilon}( z' , {\bf c}), \quad \mbox{whenever} \quad z\neq z'.
\end{equation}

Let now $\{z (t)\}_{t \in \Bbb Z} $, $\{z' (t)\}_{t \in \Bbb Z} $ be two input signals that are identical except at $s \in \mathbb{N} $, that is, $z (s) \neq z ' (s) $. We show by induction that the corresponding reservoir outputs $\mathbf{x} (t) $, $\mathbf{x} ' (t) $ are such that  $\mathbf{x} (t) \neq \mathbf{x} ' (t) $ for any $t \ge s$. First, since $z (t) = z ' (t) $ for any $t < s$, then $\mathbf{x} (t) = \mathbf{x}' (t) $ for any $t<s$ necessarily. Now by \eqref{linearization}  and using the notation introduced in~\eqref{notation epsilon}, we have:
\begin{equation*}
\mathbf{x} (s) - \mathbf{x} '(s) = A(\mathbf{x} _0 , \boldsymbol{\theta} ) (\mathbf{x} (s-1) - \mathbf{x} ' (s-1)) + \boldsymbol{\varepsilon }( z (s), C) - \boldsymbol{\varepsilon }( z '(s), C) = \boldsymbol{\varepsilon }( z (s), C) - \boldsymbol{\varepsilon }( z' (s), C).
\end{equation*}
Given that $z (s) \neq z '(s)$, the hypotheses {\bf (a)} or {\bf (b)} in the statement of the theorem together with~\eqref{injective epsilon} imply that $\mathbf{x} (s) \neq \mathbf{x} '(s)$. We now establish the induction step by assuming that $\mathbf{x} (t) \neq \mathbf{x} ' (t) $ for some $t\ge s $ and we show that $ \mathbf{x} (t+1) \neq \mathbf{x} ' (t+1)$. Indeed, in this case 
\begin{equation}
\label{}
\mathbf{x} (t+1) - \mathbf{x} '(t+1) = A(\mathbf{x} _0 , \boldsymbol{\theta} ) (\mathbf{x} (t) - \mathbf{x} ' (t)).
\end{equation}
Since by hypothesis zero is not an eigenvalue of $A(\mathbf{x} _0 , \boldsymbol{\theta} )$, we have that $\mathbf{x} (t+1) \neq \mathbf{x} ' (t+1)$ necessarily because $\mathbf{x} (t) - \mathbf{x} ' (t) \neq 0$ by the induction hypothesis.

\medskip

\noindent {\bf Proof of part {\bf  (ii)}} Let $\varepsilon _1 >0$ and $h \in \mathbb{N} $. The continuity of the map $ \boldsymbol{\varepsilon}( \cdot , {\bf c}) $ defined in~\eqref{notation epsilon} implies that there exists $\delta ( \varepsilon _1 ) >0$ such that  if  two input signals $ \{z (t)\}_{t \in \Bbb Z} $, $\{z ' (t)\}_{t \in \Bbb Z} $ are such that  $|z (s) - z' (s)|< \delta (\varepsilon _1 )$ for all $s \in \left[ t-h, t\right] $, then $\|\boldsymbol{\varepsilon}( z (s), {\bf c})  - \boldsymbol{\varepsilon}( z' (s), {\bf c}) \|< \varepsilon _1 $, for all $s \in \left[ t-h, t\right] $. Additionally, since by hypotheses the input signals are  bounded, then $\|\boldsymbol{\varepsilon}( z(t), {\bf c}) \|< K_{{\rm max}} $ and $\|\boldsymbol{\varepsilon}( z'(t), {\bf c}) \|< K_{{\rm max}} $,  for all $t \in \Bbb Z  $ and for some $K_{\rm max} \in \mathbb{R} $.

Let now $\mathbf{x} (t) $ and $\mathbf{x} ' (t) $ be the reservoir outputs corresponding to $ z (t) $ and $ z ' (t) $, respectively, then 
\begin{equation*}
\label{}
\mathbf{x} (t) - \mathbf{x} ' (t) = A( \mathbf{x} _0 , \boldsymbol{\theta} ) ( \mathbf{x} (t-1) - \mathbf{x} ' (t-1)) + \boldsymbol{\varepsilon}( z (t), {\bf c}) - \boldsymbol{\varepsilon}( z' (t), {\bf c}).
\end{equation*}
Since by hypothesis $\|A( \mathbf{x} _0 , \boldsymbol{\theta} )\|<1 $ and the spectral radius $\rho (A( \mathbf{x} _0 , \boldsymbol{\theta} )$ satisfies that $\rho (A( \mathbf{x} _0 , \boldsymbol{\theta} )<\|A( \mathbf{x} _0 , \boldsymbol{\theta} )\|$ for any matrix norm, we have that  $\rho (A( \mathbf{x} _0 , \boldsymbol{\theta} )<1$. Additionally, since the input signals are strictly stationary with finite automoments up to order $2R $, we can use Proposition~\ref{stationarity of approximate reservoir} to rewrite this expression as 
\begin{equation}
\label{}
\mathbf{x} (t) - \mathbf{x} ' (t) = \sum^{\infty}_{i = 0} A( \mathbf{x} _0 , \boldsymbol{\theta} ) ^i (\boldsymbol{\varepsilon}( z (t - i), {\bf c}) - \boldsymbol{\varepsilon}( z' (t - i), {\bf c}))
\end{equation}
which implies that
\begin{align*}
\label{}
||\mathbf{x} (t) - \mathbf{x} ' (t)|| \le &\sum^{h}_{i = 0} ||A( \mathbf{x} _0 , \boldsymbol{\theta} )||  ^i ||\boldsymbol{\varepsilon} ({z} (t-i), {\bf c})- \boldsymbol{\varepsilon} ({z}' (t-i), {\bf c}) ||  + 2 K_{\rm max} \sum^{\infty}_{i = h+1} ||A( \mathbf{x} _0 , \boldsymbol{\theta} )||  ^i\nonumber\\
\le &\varepsilon _1 \sum^{h}_{i = 0} ||A( \mathbf{x} _0 , \boldsymbol{\theta} )||  ^i + 2 K_{\rm max} \dfrac{||A( \mathbf{x} _0 , \boldsymbol{\theta} )||  ^{h+1}}{1-||A( \mathbf{x} _0 , \boldsymbol{\theta} )||  }=\dfrac{\varepsilon _1 + (2 K_{\rm max}  - \varepsilon _1 )||A( \mathbf{x} _0 , \boldsymbol{\theta} )||  ^{h+1}}{1-||A( \mathbf{x} _0 , \boldsymbol{\theta} )||  }= :\varepsilon .
\end{align*}
Finally, the hypothesis $\|A( \mathbf{x} _0 , \boldsymbol{\theta} )\|<1 $ implies that $ \lim_{h \rightarrow \infty} ||A( \mathbf{x} _0 , \boldsymbol{\theta} )||  ^{h+1} =0$ and hence since $\varepsilon _1 >0$ can be made as small as desired, so is $\varepsilon $, which proves the statement. $\blacksquare$

\subsection{Proof of Corollary~\ref{corollary sp ufm properties}}
\noindent {\bf  Proof of part (i)} In order to obtain this result as a corollary of part {\bf (i)} in Theorem~\ref{theorem sp ufm properties} we need to prove the following two statements. First, that the injectivity hypotheses on $f_I^R( \cdot , {x} _0, \boldsymbol{\theta} )$ imply those about the corresponding map $F_I^R( \cdot , \mathbf{x} _0, \boldsymbol{\theta} )$ in Theorem~\ref{theorem sp ufm properties}. Second, in this case the linear map $ A(\mathbf{x} _0,  \boldsymbol{\theta} ) :=D_{\mathbf{x} } F(\mathbf{x} _0,{\bf 0} _N, \boldsymbol{ \theta }): \mathbb{R}^N \rightarrow  \mathbb{R}^N$ has no zero eigenvalues.

The first statement can be proved by noting that, as a consequence of~\eqref{discretized reservoir}:
\begin{equation}
\label{I map}
F_I^R( \mathbf{I} , \mathbf{x} _0, \boldsymbol{\theta} )= (1- e^{- \xi })  \left( \begin{array}{c} f_I^R( {I}_1, {x} _0, \boldsymbol{\theta}   ) \\ e^{- \xi }f_I^R( {I}_1, {x} _0, \boldsymbol{\theta}    )+f_I^R( {I} _2, {x} _0, \boldsymbol{\theta}   ) \\e^{-2 \xi }f_I^R( {I}_1, {x} _0, \boldsymbol{\theta}  )+e^{- \xi }f_I^R( {I} _2 , {x} _0, \boldsymbol{\theta}  )+f_I^R( {I}_3  , {x} _0, \boldsymbol{\theta}  )\\\vdots \\ e^{-(N-1) \xi }f_I^R( {I}_1, {x} _0, \boldsymbol{\theta}   )+e^{-(N-2) \xi }f_I^R( {I} _2 , {x} _0, \boldsymbol{\theta} )+\cdots+f_I^R( {I}_N , {x} _0, \boldsymbol{\theta}  ) \end{array}\right).
\end{equation}
This expression can be used to  to easily show recursively that $F_I^R( \cdot  , \mathbf{x} _0, \boldsymbol{\theta} ) $ is injective if $f_I ^R( \cdot  , x _0, \boldsymbol{\theta} )$ is injective.  Concerning the second point,
Theorem D.11 in \cite{GHLO2014_capacity} proves that zero is not an eigenvalue of $A(\mathbf{x} _0 , \boldsymbol{\theta} )$.

\medskip

\noindent {\bf Proof of part (ii)} It follows from part {\bf (ii)} in Theorem~\ref{theorem sp ufm properties} by noting that $|\partial_x f(x _0 , 0, \boldsymbol{\theta} )| <1$ implies (see the proof of Theorem D.10 in Supplementary Material of \cite{GHLO2014_capacity}) that 
\begin{equation*}
||A(\mathbf{x} _0 , \boldsymbol{\theta} )||_{\infty}<1. \quad \blacksquare
\end{equation*}

\addcontentsline{toc}{section}{Bibliography}
\bibliographystyle{wmaainf}

\end{document}